\documentclass[runningheads]{llncs}
\usepackage[usenames,dvipsnames]{color} 
\usepackage{stfloats}
\usepackage{xcolor}
\usepackage[super]{nth}
\usepackage{mathtools}
\usepackage{wrapfig}
\usepackage{graphicx}
\usepackage{amsmath}
\usepackage{amssymb}
\usepackage{lipsum}
\usepackage[hyphens]{url}
\usepackage{hyperref}
\hypersetup{breaklinks=true}
\usepackage[multiple]{footmisc}

\makeatletter
\newcommand{\printfnsymbol}[1]{%
  \textsuperscript{\@fnsymbol{#1}}%
}
\makeatother

\begin{document}
\title{Sub 8-Bit Quantization of Streaming Keyword Spotting Models for Embedded Chipsets}

\titlerunning{Sub 8-Bit Quantization of KWS Models}

\author{{Lu Zeng\printfnsymbol{1}, Sree Hari Krishnan Parthasarathi\thanks{Equal contribution.}, Yuzong Liu, Alex Escott, Santosh Kumar Cheekatmalla, Nikko Strom, Shiv Vitaladevuni}}

\authorrunning{Zeng, Parthasarathi et al.}
\institute{Alexa, Amazon, USA\\
\email{\{luzeng, sparta, liuyuzon, escottal, cheekatm, \\ nikko, shivnaga\}@amazon.com}}

\maketitle
\begin{abstract}
We propose a novel 2-stage sub 8-bit quantization aware training algorithm for all components of a 250K parameter feedforward, streaming, state-free keyword spotting model. For the \nth{1}-stage, we adapt a recently proposed quantization technique using a non-linear transformation with $tanh(.)$ on dense layer weights. In the \nth{2}-stage, we use linear quantization methods on the rest of the network, including other parameters (bias, gain, batchnorm), inputs, and activations. We conduct large scale experiments, training on 26,000 hours of de-identified production, far-field and near-field audio data (evaluating on 4,000 hours of data). We organize our results in two embedded chipset settings: a) with commodity ARM NEON instruction set and 8-bit containers, we present accuracy, CPU, and memory results using sub 8-bit weights (4, 5, 8-bit) and 8-bit quantization of rest of the network; b) with off-the-shelf neural network accelerators, for a range of weight bit widths (1 and 5-bit), while presenting accuracy results, we project reduction in memory utilization. In both configurations, our results show that the proposed algorithm can achieve: a) parity with a full floating point model's operating point on a detection error tradeoff (DET) curve in terms of false detection rate (FDR) at false rejection rate (FRR); b) significant reduction in compute and memory, yielding up to 3 times improvement in CPU consumption and more than 4 times improvement in memory consumption.
\end{abstract}


\section{Introduction}
\label{sec:intro}
Wakeword detection, also known as keyword spotting (KWS), detects words or phrases of interest from streaming audio and plays a vital role in voice assistants~\cite{chen2014small,panchapagesan2016multi}. KWS models are based on neural network architectures and are processed on device. A challenge for KWS systems is to attain high accuracy under tight resource constraints such as model size, runtime memory footprint, and power consumption. To address some of the constraints, previous work has explored knowledge distillation~\cite{tucker2016model}, low rank approximation~\cite{prabhavalkar2016compression,sun2017compressed}, and computationally efficient architectures~\cite{mittermaier2020small,li2020small,blouw2020hardware}. As an orthogonal direction, quantization has been applied to KWS, where the components of the model are converted from 32-bit floating points to lower bit width representations. 

Quantization is often applied post-training, resulting in performance degradation~\cite{vandersteegen2021integer}. To mitigate this, quantization aware training (QAT) is applied to reduce errors~\cite{shi2019compression,mishchenko2019low}. QAT, including very low bit width (even binary quantization), is an established technique~\cite{rastegari2016xnor,courbariaux2015binaryconnect}. Since the quantization function is discrete (and therefore the gradient is zero almost everywhere), a fake quantizer with straight through estimator (STE)~\cite{bengio2013estimating} or a Gumbel-Softmax trick~\cite{jang2016categorical} are used to estimate the parameters. Since accuracy of 1-bit models is still a challenge, in the context of speech processing, 4 to 6-bit QAT has been studied for event detection~\cite{shi2019compression}, speech recognition~\cite{nguyen2020quantization}, and KWS~\cite{mishchenko2019low}. 

Furthermore, commercially realizable KWS models have additional challenges often ignored in research focused on ``model size vs accuracy'' tradeoff: a) sub 8-bit quantization requires hardware support\footnote{An instruction set that can efficiently carry out matrix-vector multiplications.}; b) need to run in streaming mode\footnote{Models have to run with low latency -- i.e., cannot use large buffers.}; c) cannot use corruptible memory\footnote{Since the models are running continuously, they cannot get into a ``bad'' state.}. Our paper happens in the context of: a) using sub 8-bit representations to avoid overflow errors with 8-bit containers in commodity or off-the-shelf platforms; b) addressing accuracy challenges with sub 8-bit, including 1-bit representations.

\noindent{\textbf{Contributions:}} In this paper, we tackle the problem of sub 8-bit quantization of on-device, small footprint, streaming, state-free KWS models, that can execute on commodity or off-the-shelf hardware platforms. We propose a novel 2-stage QAT algorithm: for the \nth{1}-stage, we adapt a non-linear quantization method on weights~\cite{nikko_tanh}, while for the \nth{2}-stage, we use linear quantization methods on other components of the network. We conduct large scale experiments, training on 26K hours of de-identified production audio, collected from a mix of far-field devices and mobile phones (evaluating on 4K hours of data). We show the efficacy of our methods by presenting accuracy and compute results (CPU\footnote{We use CPU cycles as a proxy for power consumption.}, memory, and model size) for sub 8-bit models (4, 5, 8-bit) on ARM NEON chipset, while projecting memory gains for a range of weight bit widths (1 and 5-bit) on off-the-shelf neural network accelerators.

\section{Small Footprint, Streaming, State-Free KWS Models}
\label{sec:model_setup}
While QAT algorithm can be applied to convolutional models, we use a model that is a feedforward network (FFN) with 250k parameters using a bottleneck architecture~\cite{gao2020front}. The full precision version of this model has been optimized for low power, small footprint, streaming and state-free setting and serves as our baseline. It is trained on a set of positive and negative examples (i.e. positive examples contain the wakeword, while negative examples do not). 

The model architecture has 250K learnable parameters, and it operates on 20-dimensional log mel filter bank energy (LFBE) features, computed with an analysis window size and shift of 25 ms and 10 ms respectively; the input to the model is 81 frames, downsampled by a factor of 3. The architecture consists of five fully connected layers with batch normalization~\cite{ioffe2015batch} and ReLU~\cite{agarap2018deep} being used with all hidden layers. The output is a binary classification layer trained with cross entropy loss, representing the posterior probability of “wakeword” and “non-wakeword” ~\cite{gao2020front}. Later, to investigate quantized models with lower bit width, we explore two other model sizes with different layer sizes, keeping the other architectural aspects the same. During training, the wakeword is consistently center aligned in the input window~\cite{Jose}. Adam optimizer is used to update the model parameters during training. During inference, the posterior estimates corresponding to the wakeword are smoothed by a windowed smoothing average (WMA) filter and then thresholded to infer the wakeword hypothesis.

\section{Relevant Embedded Chipsets}
\label{sec:arm_neon}
In this section, we describe hardware considerations in two settings: a) commodity ARM NEON instruction set; b) off-the-shelf neural network accelerators.

\noindent\textbf{{Commodity ARM NEON Instruction Set Overview:}}
In 2021 over 90\% of mobile phones use ARM-based chipsets~\footnote{https://www.counterpointresearch.com/global-smartphone-ap-market-share/}. The majority of these include a Single-Instruction-Multiple-Data (SIMD) extension known as NEON; this is available by default in ARM's chipsets for mobile phones (\textit{armv8a}, \textit{aarch64}~\footnote{https://www.arm.com/blogs/blueprint/android-64bit-future-mobile}). The NEON instruction set is based on parallelizing arithmetic operations on vectors stored in 128-bit registers, allowing packing of multiple scalar values in the register – i.e. 4x32-bit floating point, 4x32-bit integer, 8x16 bit-integer, or 16x8-bit integer – then perform an arithmetic operation on every value in the register in a clock cycle. For NEON, 8-bit is the smallest supported container for Multiply-And-Accumulate (MAC) operations.

The compute requirement for a KWS model can be reduced by using lower bit widths for weights, input and and activations. Our QAT algorithm produces a model that can utilize the SIMD instruction set available with NEON. While NEON uses 8-bit containers for MAC operations, using sub 8-bit weights can yield benefits: a) prevent overflow when performing computations; b) lower model size for over the network deployments.

\noindent\textbf{{Off-the-shelf Neural Network Accelerators:}}
The smallest container in ARM NEON architecture is 8-bit, so we do not observe a memory reduction with sub 8-bit weights. Some off-the-shelf accelerators can utilize sub 8-bit weights~\footnote{https://datasheets.maximintegrated.com/en/ds/MAX78000.pdf}\footnote{https://www.syntiant.com/post/syntiant-introduces-second-generation-ndp120-deep-learning-processor-for-audio-and-sensor-apps}. In this paper, we use data sheets from them to project memory savings for our QAT algorithm.

\section{Proposed 2-Stage QAT Algorithm}
\label{sec:e2e_method_description}
In this section, we describe the design of our approach. Sections~\ref {sec:overall_method} gives an overview of our 2-stage training algorithm. We provide detailed descriptions of the first and second stages in Sections~\ref{sec:nikko_method} and~\ref{sec:second_stage}.

\subsection{Quantization Overview}
\label{sec:overall_method}
A layer in a feedforward neural network typically takes the following form: 
\begin{equation}
    \mathbf{y} = \psi_{bn}(\phi_{relu}(\mathbf{W}\mathbf{x}+\mathbf{b}))
\end{equation}
where $\mathbf{x}$ is the input to the dense layer, $\mathbf{W}$ is the dense layer weight matrix,  $\mathbf{b}$ is the bias, $\phi_{relu}$ and $\psi_{bn}$ represent ReLU and batch norm transformations. Using a gain $\mathbf{\alpha}$ and a linear quantizer ($q_z(.)$) that discretizes the values to $z$ levels ($\log_{2}(z)$ bits), let $\mathbf{W_q}$, $\mathbf{x_q}$, $\mathbf{\alpha_q}$, $\mathbf{b_q}$ be the corresponding quantized representations. Quantization of such a layer can be done in the following steps: 
\begin{gather*}
\label{eq:dense_layer}
	\mathbf{y_q} \leftarrow \mathbf{\alpha_q} (\mathbf{W_q} \cdot \mathbf{x_q} +\mathbf{b_q}) \\
	\mathbf{y_q^{(\phi_q)}} \leftarrow q_z(\phi_{relu} (\mathbf{y_q})) \\
	\mathbf{y_q^{{(\phi_q}, \psi_{q})}} \leftarrow q_z(\psi_{bn} (\mathbf{y_q^{(\phi_q)}}))
\end{gather*}
Our QAT algorithm for the layer follows a 2-stage procedure. During the \nth{1}-stage we adapt the $tanh(.)$ based quantization technique introduced in~\cite{nikko_tanh}, for obtaining $\mathbf{W_q}$. In the \nth{2}-stage of training, we propose specific linear quantization techniques for the remaining components: gain ($\mathbf{\alpha_q}$), bias ($\mathbf{b_q}$), batch normalization ($q_z(\psi_{bn}(.))$), activations ($q_z(\phi_{relu}(.))$), and inputs ($\mathbf{x_q}$).

\subsection{\nth{1} Stage: $tanh(.)$ Quantization of Weights}
\label{sec:nikko_method}
Linear quantization \cite{vanhoucke2011improving} is widely used for model compression. However, the weights of a neural network are typically Gaussian distributed. Linear quantization is not efficient for Gaussian distributions. To mitigate this problem, we adopt~\cite{nikko_tanh}: it encourages the weights of a neural network to become more uniform distributed, by first applying $tanh(.)$ on $\mathbf{W}$, and then applying a linear quantizer to obtain $\mathbf{W_q} \leftarrow q_z(\mathbf{\tanh(W)})$; here $\mathbf{W_q}$ is the quantized weight matrix.

\begin{figure}[ht]
  \centering
  \includegraphics[width=0.8\linewidth]{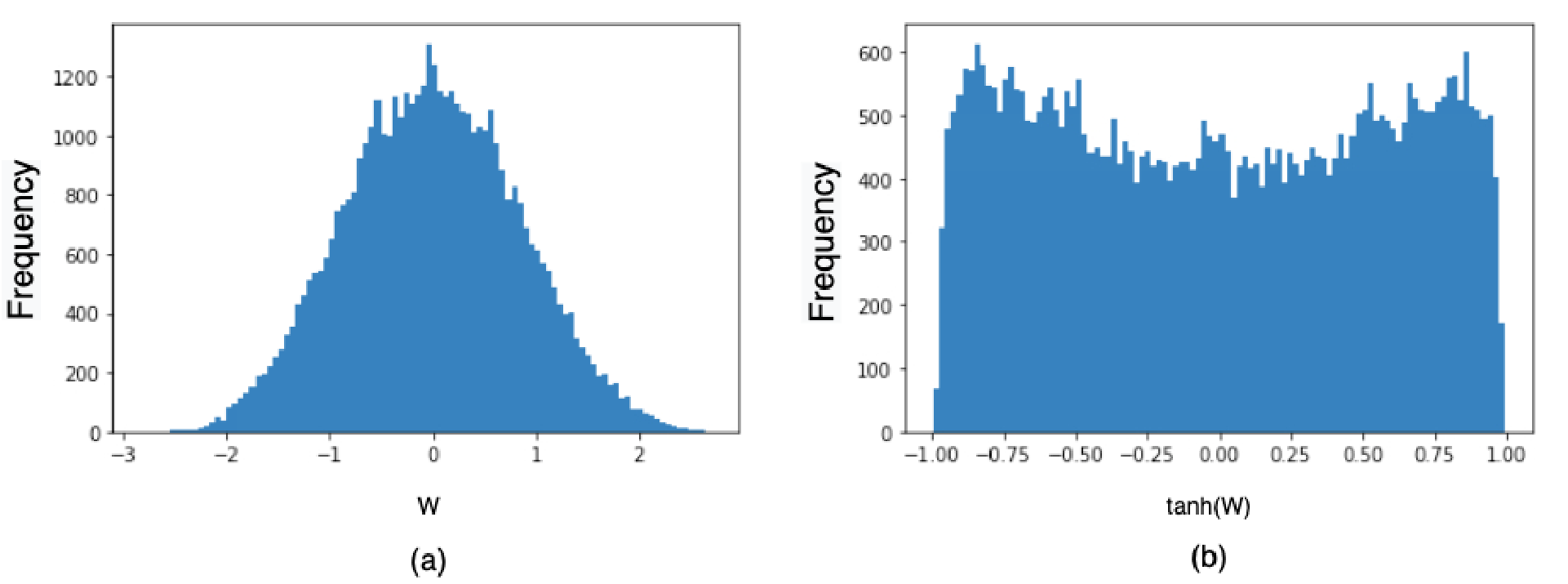}
  \caption{\textit{Distributions for (a) original dense layer weights with a nearly Gaussian distribution; (b) dense layer weights with a flatter distribution after $tanh(.)$.}}
 \label{fig:kernel_distri}
\end{figure}

Fig. \ref{fig:kernel_distri}(a) shows an example on the original weight distribution $\mathbf{W} \sim \mathcal{N}(0, \sigma_{w}^2)$. In order to push the weight distribution towards $\mathcal{U}(-1,1)$, the weights are initialized with a target distribution $\mathcal{N}(0,\sigma_{t}^2)$ and a regularizer $\lambda_{\sigma} (\sigma_{w} - \sigma_{t})^2 + \lambda_{\mu} \mu_{w}^2$ is added during training to penalize deviations from the target weight distribution, where $\sigma_{w}$ and $\mu_{w}$ are the standard deviation and mean of the weights. An example of the resultant distribution is shown in Fig. \ref{fig:kernel_distri}(b).

\subsection{\nth{2} Stage: Linear Quantization of Full Network}
\label{sec:second_stage}
In this section we describe the second stage, providing details on quantization of gain and bias, batch norm, activations, and lastly the inputs. We then perform one epoch of training to reestimate parameters.
\subsubsection{Quantization of BatchNorm.}
\label{sec:quant_bn}
\noindent 
Recall that BatchNorm (BN) has four sets of parameters, $\gamma$, $\beta$, $\mu_{B}$ and $\sigma^2_{B}$. While $\gamma$ and $\beta$ are trainable parameters, $\mu_{B}$ and $\sigma^2_{B}$ are the empirical estimates of means and variances. These are running estimates of mean and standard deviation during training, while they are replaced by global estimates during inference. In this work, we quantize all four sets of parameters.

From the \nth{1}-stage, we inherit floating point representations of the four sets of parameters. During this stage of training (i.e., \nth{2}), we quantize these parameters. To quantize the BN parameters, given a preferred bit-width, we apply linear quantization on $\gamma$, $\beta$, $\mu_{B}$ and $\sigma^2_{B}$ separately. Following \cite{banner2018scalable}, BN parameters, especially $\mu_{B}$ and $\sigma^2_{B}$, tend to have large dynamic range. We also observe that the parameters continuously shift during training. We start with $\mu_{B}$ and $\sigma^2_{B}$: to reduce the dynamic range, we introduce a scale factor $C_{BN}$ and quantize $\mathbf{x}$, $\mu_{B}$ and $\sigma^2_{B}$ with a linear quantizer ($q_z(.)$):
\begin{equation}
\label{lab:eq_norm}
    \begin{aligned}
 	\mathbf{x^{norm}} = \frac{\mathbf{x}- \mu_{B}}{\sqrt{\sigma^2_{B}}} = \frac{\frac{\mathbf{x}}{C_{BN}}-\frac{\mu_{B}}{C_{BN}}}{\sqrt{\frac{\sigma^2_{B}}{C_{BN}^2}}} \\
 	\mathbf{x^{norm}_q} \leftarrow \frac{q_z(\frac{\mathbf{x}}{C_{BN}})-q_z(\frac{\mu_{B}}{C_{BN}})}{\sqrt{q_z(\frac{\sigma^2_{B}}{C_{BN}^2})}} \\
 	\end{aligned}
\end{equation}
We then perform the remaining steps in quantizating a BN transformation, with the application of the $q_z(.)$ on $\gamma$ and $\beta$ to obtain $q_z(\psi_{bn} (\mathbf{x}))$:
\noindent
\begin{gather*}
    \gamma_q \leftarrow q_z(\gamma) \\
    \beta_q \leftarrow q_z(\beta) \\
 	q_z(\psi_{bn} (\mathbf{x^{norm}_q})) \leftarrow \gamma_q \cdot \mathbf{x^{norm}_q} + \beta_q
 \end{gather*}

\subsubsection{Quantization of Gain and Bias.} 
\label{sec:quant_gain_bias}
We apply a linear quantizer on gain ($\mathbf{\alpha}$) and bias ($\mathbf{b}$) to obtain the respective quantized representations $\mathbf{b_q}$ and $\mathbf{\alpha_q}$.
\subsubsection{Quantization of Activation Functions.}
A standard dense layer is typically followed by a non-linear activation function and a BN layer in order. Our initial experiments showed that a direct linear quanization of ReLU can lead to a large drop in accuracy. To mitigate this, since ReLU is unbounded on the positive domain, we experimented with bounded (clipped ReLU, sigmoid, and tanh) as well as with smoother activation functions (GeLU, SiLU)\cite{hendrycks2016gaussian,Elfwing2018SigmoidWeightedLU}. Clipped ReLU was the most promising activation function in our initial study; experiments with GeLU and SiLU did not yield conclusive results; while sigmoid and tanh yielded worse performance than with a direct quantization of ReLU (presumably require more training updates due to gradient saturation). The output of clipped ReLU is processed with the linear quantizer to obtain $q_z(\phi_{relu}(.))$. We also found that with quantization, the order of BN and the activation matters. Specifically, we switch the order of BN and clipped ReLU activations. 

\subsubsection{Quantization of Inputs.} 
The input LFBE features are processed using global mean and variance normalization. Subsequent to this step, the normalized input is processed similar to Section~\ref{sec:quant_bn}, such that $\mathbf{x^{input}_q} \leftarrow q_z(\frac{\mathbf{x^{input}}}{C_{input}})$.

\section{Experimental Setup}
\label{sec:experimetal_setup}
In this section we describe our training and test datasets; we also discuss the models and evaluation metrics. All experiments in this paper were conducted on de-identified production datasets. 

\noindent\textbf{Datasets:} For our experiments we used a fully labeled training dataset consisted of 26K hours of audio. The training dataset contains both far-field audio and near-field mobile phone audio. We used two test sets in this work: (a) a validation test set (referred to as VAL), which consisted of about 4K hours of audio data, (b) an independent test set (referred to as TEST), which also consisted of about 4K hours of audio data. Both VAL and TEST data contain far-field audio and near-field mobile phone audio. TEST data was collected from a wider range of commodity devices.

\noindent\textbf{Evaluation metrics:} During inference, we tuned the WMA values for the models on held-out datasets. We measure the model performance with DET curves having False Rejection Rate (FRR) on the x-axis and False Discovery Rate (FDR) on the y-axis. Similar to ~\cite{gao2020front,Sun,Jose}, we normalize the axes of DET curves and report relative FDR. In the interest of space, we do not present the full DET curves for all experiments; in such cases we only report relative degradation in FDR at the FRR of the baseline model's operating point (OP), where the chosen OP corresponds to the $1.0$ in relative FRR.

\noindent\textbf{Models: } All our models use the architecture described in Section~\ref{sec:model_setup}. The proposed QAT technique, presented in Section~\ref{sec:e2e_method_description}, is applied to different components of the model, and the results are described in Section~\ref{sec:results_e2e}. We also study the effect of number of bits for parameters (weights, biases, gain, BN parameters), input and activations. Details of the models are summarized in Table~\ref{tab:ID_architecture}. M0 is a full precision model. M1 uses 8-bit parameters and 16-bit activation and input. M2, M3 and M4 use weights quantized to 8, 5, and 4 bits respectively, while the activations and input are quantized to 8-bits. Models M0 to M4 have 250K learnable parameters.

\begin{table*}[t]
  \caption{\textit{Model architecture including hidden layer size, \# param., \# bits for weights, \# bits for gain ($\alpha$), bias ($b$), and BN parameters ($\mu, \sigma, \gamma, \beta$), and input and activation.}}
  \label{tab:ID_architecture}
  \centering
  \begin{tabular}{c|c|c|c|c|c|c}
    \multicolumn{1}{c}{\textbf{\shortstack{ID}}}  & 
    \multicolumn{1}{c}{\textbf{\shortstack{Layer Size}}} &
    \multicolumn{1}{c}{\textbf{\shortstack{\# Param.}}} & 
    \multicolumn{1}{c}{\textbf{\shortstack{\# bits \\weight}}} & 
    \multicolumn{1}{c}{\textbf{\shortstack{\# bits \\$(\alpha, b, BN)$}}}& 
    \multicolumn{1}{c}{\textbf{\shortstack{\# bits \\act.}}} &
    \multicolumn{1}{c}{\textbf{\shortstack{\# bits \\input}}}\\
    \hline
    M0 &   
    \{87, 400, 87, 400, 87, 400\} & 
    250K &
    32& 32& 32 & 32 \\
    \hline
    M1 &   
    \{87, 400, 87, 400, 87, 400\} &  
    250K &
    8& 8& 16& 16 \\
    M2 &  
    \{87, 400, 87, 400, 87, 400\} &  
    250K &
    8& 8& 8& 8 \\
    M3 &  
    \{87, 400, 87, 400, 87, 400\} &  
    250K &
    5& 8& 8& 8\\
    M4 &   
    \{87, 400, 87, 400, 87, 400\} &  
    250K &
    4& 8& 8& 8\\
    \hline
    \end{tabular}
\end{table*}

\section{Results}
\label{sec:results_e2e}
In this section, we provide a detailed study of the proposed 2-stage QAT approach, against an unquantized full precision model in terms of (a) accuracy, (b) memory and computation. Our results are presented in 2 groups: a) Non-binary, sub 8-bit models in Sections~\ref{sec:effective_tanh} and~\ref{sec:model_performance_full_quant}, for ARM NEON instruction set; b) Binary (1-bit) weight models in Section~\ref{sec:binary_1bit_2nd_stage}, for off-the-shelf accelerators.

\subsection{Non-binary Sub-8 bit Models: \nth{1}-Stage Training}
\label{sec:effective_tanh}
To study the effectiveness of $tanh(.)$, we train models with 5-bit dense layer weights for 100K updates, with and without $tanh$; note that other components are not quantized in this experiment, and that all trainable parameters are updated. Table~\ref{tab:tanh_quant} presents the results, comparing against an unquantized baseline model. The table also presents results with two quantizers for weights: a) the non-linear quantization with $tanh(.)$; b) a linear quantizer without $tanh(.)$. The model with $tanh(.)$ achieves a 7.4\% relative degradation in FDR, while the model without $tanh(.)$ yields a 12.3 \% degradation in FDR.

\begin{table}[h]
\caption{\label{tab:tanh_quant} \textit{Relative degradation in FDR at baseline model's FRR on VAL. Models have 5-bit weights with or without $tanh(.)$.}}
  \centering
 \begin{tabular}{l|c}
    \multicolumn{1}{c}{\textbf{\shortstack{Quantization}}} & \multicolumn{1}{c}{\textbf{\shortstack{Rel. FDR ($\%$)}}} \\
    \hline
   Unquantized model     & \shortstack[c]{0.0 \\(Baseline)}   \\
   \hline
   With $tanh(.)$     & $7.4$  \\
   Without $tanh(.)$    & $12.3$    \\
   \hline
  \end{tabular}
\end{table}

\subsection{Non-binary Sub-8 bit Models: \nth{2}-Stage Training}
\label{sec:model_performance_full_quant}
In this section, we compare the performance of the quantized models against the baseline unquantized model. The baseline model is trained for 500K model updates. For the proposed QAT approach, the models were trained for 500K and 35K updates in the \nth{1} and \nth{2}-stage training respectively. Firstly, we discuss the accuracy implications, and then present results in terms of memory and CPU gains (for ARM NEON instruction set).

\subsubsection{Evaluation of Accuracy.}
From Table~\ref{tab:ID_architecture_performance}, M1 with the proposed 2-stage training algorithm yields a $3.1\%$ relative degradation in FDR. To study the effect of number of bits for weights, we quantized M2, M3 and M4 with 2-stage training. Notice that going from M0 to M1, there is a very small increase in relative degradation in FDR at FRR (3.1\% increase). Here the changes include: 32 to 8-bit weights; 32 to 8-bit quantization of other parameters (bias, gain, BN parameters); and 32 to 16-bit input and activations. Further reducing the inputs and activations from 16-bit to 8-bit (i.e., M1 to M2) does not lead to an increase in FDR (3.1\% to 3.5\%). A further change in weights from 8 to 5 or 4-bits (i.e., M2 to M3 or M4) leads to a small increase in FDR (3.5\% to 7.4\% or 8.7\%). 

\begin{table*}[b]
  \caption{\textit{Performance of models in terms of rel. degradation in FDR at FRR against baseline full precision model on TEST data. CPU, memory consumption, and model size for ARM NEON in MCPS and KB (MCPS refers to Million Cycles Per Second).}}
  \label{tab:ID_architecture_performance}
  \centering
 \begin{tabular}{c|c|c|c|c|c|c|c|c}
    \multicolumn{1}{c}{\textbf{\shortstack{ID}}}  & 
    \multicolumn{1}{c}{\textbf{\shortstack{\# bits \\weight}}} & 
    \multicolumn{1}{c}{\textbf{\shortstack{\# bits \\act.}}} &
    \multicolumn{1}{c}{\textbf{\shortstack{\# bits \\input}}} &
 \multicolumn{1}{c}{\textbf{\shortstack{Quant. \\Method}}} & 
    \multicolumn{1}{c}{\textbf{\shortstack{Rel. FDR \\Degrad. ($\%$)}}} &
    \multicolumn{1}{c}{\textbf{\shortstack{CPU \\ (MCPS)}}} & 
    \multicolumn{1}{c}{\textbf{\shortstack{Memory \\(kb) }}} &
    \multicolumn{1}{c}{\textbf{\shortstack{Model \\size (kb) }}} \\
    \hline
    M0 & 
    32 &
    32 &
    32 &
    No Quant. &
    \shortstack[c]{0.0 \\(Baseline)} &
    37.6 &
    912 &
    912\\
    \hline
    M1 &
    8 &
    16 &
    16 &
    2-stage &  
    3.1 &
    18.8 &
    228 &
    228\\
    M2 &
    8 &
    8 &
    8 &
    2-stage &  
    3.5 &
    10.3 &
    228 &
    228\\
    M3 & 
    5 &
    8 &
    8 &
    2-stage &  
    7.4&
    10.3 &
    228 &
    142\\
    M4 &  
    4 &
    8 &
    8 &
    2-stage &  
    8.7 &
    10.3 &
    228 &
    114\\
     \hline
    \end{tabular}
\end{table*}

\subsubsection{CPU, Memory, and Model Size Gains.}
Reduction in compute resources (memory, CPU) is dependent on hardware architecture. In Table~\ref{tab:ID_architecture_performance}, we present CPU, memory consumption on ARM NEON.

\noindent\textbf{CPU:} We observe a 45\% reduction in CPU when using 8-bit activation and input versus 16-bit activation and input. This reduction in bandwidth is due to 8-bit SIMD processors achieving a throughput of 2x computation with 8x8 bit multiplications compared to 8x16 bit multiplications. Note that going from a full precision M0 model to M1 (a 8-bit parameters, 16-bit activations and input), lead to a $50\%$ saving in CPU consumption. Also, since 8-bit containers are used, further reduction in bit widths (i.e. from M2 to M4), do not yield gains in MCPS.

\begin{table}
\caption{\label{tab:other_parameter_and_act_quantization}\textit{Relative degradation in FDR at FRR against baseline on TEST, compressing weights, other parameters, input and activations to 8-bit.}}
  \centering
 \begin{tabular}{lc}
    \multicolumn{1}{c}{\textbf{\shortstack{Quantization on}}} & \multicolumn{1}{c}{\textbf{\shortstack{Rel. FDR Degrad. ($\%$)}}}\\
    \hline
   weight     & $3.8$  \\
   weight + bias + gain     & $4.7$    \\
   weight + input + activation     & $5.2$    \\
   weight + $BN_1$      & $5.3$    \\
   weight + $BN_2$      & $9.0$    \\
  \end{tabular}
\end{table}

\noindent\textbf{Memory and Model size:} We observe a 75\% reduction in memory by using 8-bit weights, input and activations when compared to non-quantized, 32-bit models. The smallest data type in ARM NEON is 8-bit, so we don’t observe any memory reduction by using sub-8-bit weights. However, we study and use sub 8-bit weights because this effectively provides more headroom for accumulation of weights, input and activations -- reducing the likelihood of overflow, and using a lower-precision accumulator. We also note the model size, which decreases linearly with bit width.

\subsubsection{Some Ablation Studies.}
\label{sec:other_para_q}
In this segment, we present 3 sets of ablation studies: 1) quantizing different components of the network to 8-bits; 2) quantizing inputs and activations from 8-bit to 5-bit; 3) quantizing other components of the network from 8-bit to 5-bit (bias, gain, batch norm parameters).

\begin{figure}[h]
  \centering
  \includegraphics[width=0.85\linewidth]{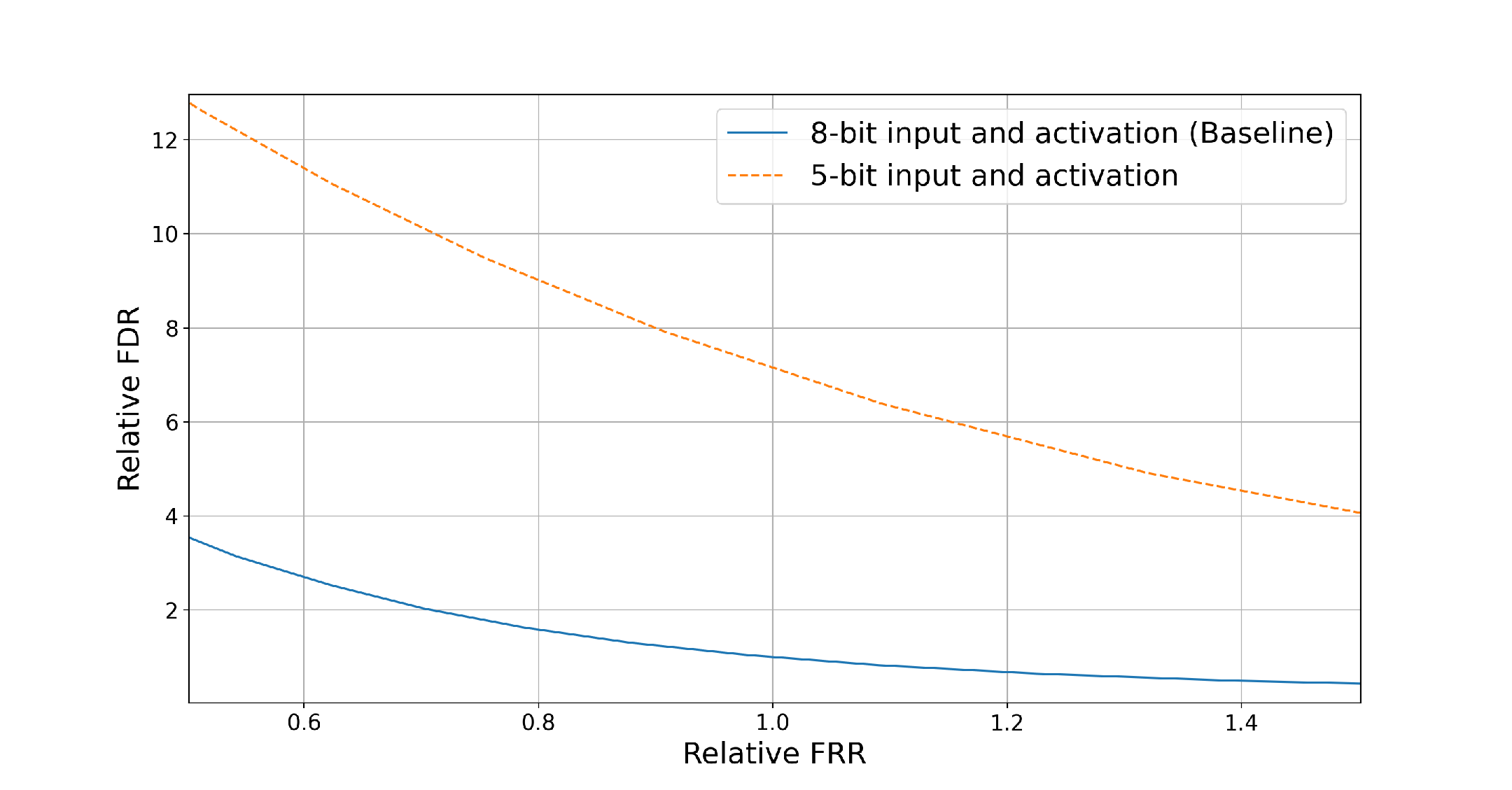}
  \caption{\label{fig:activation_bit_size}\textit{On TEST: DET curves showing performance of quantized model with 8-bit input and activation (Baseline, Blue) and 5-bit quantization (Orange)}}
\end{figure}

\noindent\textbf{1) 8-bit models}: Fixing 8-bit quantization for all components of the network, we perform an ablation study for accuracy implications in terms of quantizing bias, gain, BN parameters, input and activations. The unquantized baseline model  and the quantized models in this section were trained for 100K updates. Table~\ref{tab:other_parameter_and_act_quantization} presents the results in terms of relative degradation in FDR at equal FRR compared to the baseline model on TEST data. With $tanh(.)$ 8-bit weight quantization we obtain a 3.8 \% relative degradation in FDR. With further quantization of input and activations (5.2\%) or bias and gain (4.7\%) or batch norm parameters (5.3\%), we did not see a significant decrease in model performance. However, it is interesting to note from the Table~\ref{tab:other_parameter_and_act_quantization} that the dynamic scaling of BN yields an improvement (i.e., from 9.0\% to 5.3\% relative FDR degradation).

\noindent\textbf{2) 8-bit to 5-bit change in input and activations}: While compressing input and activations from 16-bit to 8 bits (from M1 to M2 in Table~\ref{tab:ID_architecture_performance}) only results in a minor performance degradation, reducing further from 8-bit to 5-bit input and activations leads to a significant drop in performance (see Fig. \ref{fig:activation_bit_size}).

\noindent\textbf{3) 8-bit to 5-bit change in parameters}: Similar to the small degradation in FDR in going from 8 bits to 5 bits for weights (Table~\ref{tab:ID_architecture_performance}), reducing the bit width from 8 to 5 for other parameters does not lead to a significant degradation in FDR (3.0\% increase in relative FDR).

\subsection{Binary Weight Models}
\label{sec:binary_1bit_2nd_stage}

In this section, we present results for models with 1-bit weights using 2-stage training; note that the other parameters, input and activations are non-binary (i.e., either quantized or full precision). The models were trained for 500K and 35K updates in the \nth{1} and \nth{2}-stage training respectively.

We experimented with two quantized models, with 2-times and 6-times the number of parameters as the baseline model (250K parameters): a) M5 with 500K parameters -- where the first, second, and the last layer weights are 5-bit, the other layer weights are 1-bit, and all other parameters, input and activations being 8-bit; b) M6 with 1.5M parameters -- where all layer weights are 1-bit, and other parameters, input and activations being 8-bit. Results are reported in Table~\ref{tab:binary_model_performance}. In addition to FDR, we report the model size and project reduction in memory utilization with off-the-shelf neural network accelerators.

\begin{table*}[b]
  \caption{\textit{Performance of the models in terms of rel. degradation in FDR at FRR matching baseline full precision model on TEST. Hidden layer sizes of M5 are \{123, 566, 123, 566, 123, 566\}, while those of M6 are \{1080, 566, 123, 566, 123, 566\}.}}
  \label{tab:binary_model_performance}
  \centering
 \begin{tabular}{c|c|c|c|c|l}
    \multicolumn{1}{c}{\textbf{\shortstack{ID}}}  & 
    \multicolumn{1}{c}{\textbf{\shortstack{\# Param.}}}  &
    \multicolumn{1}{c}{\textbf{\shortstack{Rel. FDR \\Degrad. ($\%$)}}} &
    \multicolumn{1}{c}{\textbf{\shortstack{Memory \\ (kb)}}} &
    \multicolumn{1}{c}{\textbf{\shortstack{Model \\size (kb) }}} &
    \multicolumn{1}{c}{\textbf{\shortstack{Note}}} \\
    \hline
    M0 & 
    250K &
    \shortstack[c]{0.0} &
    912 &
    912 & Baseline\\
    \hline
    M5 &  
    500k & 
    14.0 &
    128 &
    128 &
    with large LR\\
    M6 &  
    1.5M &
    6.2 &
    198 &
    198 &
    with large LR\\
    \hline
    M5 &  
    500k &
    0.4 &
    128 &
    128 &
    with LR change\\
    M6 &  
    1.5M &
    -0.4 &
    198 &
    198 &
    with LR change\\
     \hline
    \end{tabular}
\end{table*}

In our experiments, models having binary weights are sensitive to learning rates. Although Adam optimizer is employed in model training, we find that manually changing learning rate to a smaller number for the last few model updates of the \nth{1}-stage improves the model performance for M5 and M6. In Table~\ref{tab:binary_model_performance}, we find that the learning rate change results in 13.6\% and 6.6\% reduction in relative degradation in FDR against baseline model (M0) for M5 and M6 respectively. With the LR change, M5 and M6 are able to achieve performance parity with M0. With off-the-shelf neural network accelerators, projected memory utilization is equivalent to the model size, with 86.0\% and 78.3\% reduction in memory consumption with M5 and M6.

\section{Conclusions}
\label{sec:conclusions}

We proposed a 2-stage algorithm for sub 8-bit quantization of 250K parameter KWS models. For the \nth{1}-stage, we adapted a recently proposed QAT technique using a non-linear transformation on weights. In the \nth{2}-stage, we used linear quantization methods on other parameters, input and activations. This paper happens in the setting of on-device, low footprint, streaming KWS models, that being explored on two embedded chipset settings, where we achieved parity in accuracy against a full precision model in terms of FDR at a chosen FRR. With sub 8-bit non-binary weight models, on an ARM NEON architecture, we match accuracy, and obtain up to 3 times improvement in CPU consumption and more than 4 times improvement in memory consumption. With a binary weight model (and other components being 8-bit), using off-the-shelf neural network accelerators, at accuracy parity with a full precision model, we project 4 to 7 times reduction in memory consumption and model size.

\bibliographystyle{IEEEtran}
\bibliography{tsd1127a}

\end{document}